\def\z0{\rm Z^0}
\newcommand{\as}{\alpha_{\rm s}}

\newcommand{\oaa}{{\cal O}(\as^2)}
\newcommand{\oaaa}{{\cal O}(\as^3)}
\newcommand{\epem}{\rm e^+\rm e^-}

\newcommand{\amz}{\as(M_{\rm Z^0})}

\def\d2{D_2}
\def\oq{\char'134}

\def\ecm{E_{cm}}

\def\m2{\mu^2}
\def\q{\rm q}

\def\p{\rm p}

\def\q2{Q^2}

\documentstyle[epsf,twoside,fleqn,espcrc2]{article}
\begin{document}
 
\title{QCD TESTS AT $e^+e^-$ Colliders}
\author{Siegfried Bethke\address{III. Physikalisches Institut, RWTH, D - 52056
Aachen, Germany}%
 }

\begin{abstract}
A short review of the history and a `slide-show' 
of QCD tests in $\epem$ annihilation is given. 
The world summary of measurements of $\as$ is updated.
%
\vskip-74mm
   {\small \noindent
  Talk presented at the {\it QCD Euroconference 97}, Montpellier (France)
July 3-9, 1997.} 
   \begin{flushright} {\large PITHA 97/37} \\
    {\large September 1997}\\
    {revised: November 1997}
  \end{flushright}
\vskip50mm 
\end{abstract}
\maketitle

\section{INTRODUCTION}

Hadronic final states of $\epem$ annihilations have proven to provide 
precise tests of the strong interaction
and of its underlying theory, Quantum Chromodynamics (QCD).
The success and increasing significance of QCD tests which was achieved
in the past few years, especially with the precise data from LEP, 
was based on improvements and a deeper understanding of the
theoretical predictions as well as of the experimental techniques,
which both substantially profited from the experience of earlier
studies with data from lower energy colliders. 

In fact, in many respects it seems desirable to return to the low
energy $\epem$ data and reanalyse them with the knowledge of today. 
One reason behind this demand is that the size as well as the
energy dependence of
the strong coupling parameter $\as$ is largest at lower energies, and
thus the characteristic feature of QCD, asymptotic freedom and the
running of $\as$, can be most significantly tested if reliable data
at `low' energies are available; see \cite{qcd96} for a recent review.

In the light of these remarks, a short historical review of QCD tests
in $\epem$ annihilation will be given in this report.
In Section~2, selected highlights of early QCD studies, at energies
below the $\z0$ pole, will be reviewed,
and some of the original illustrations will be reproduced.
Section~3 contains a brief overview of QCD tests achieved at LEP and
at the SLC.
In section~4, the latest measurements of $\as$ from LEP at energies
above the $\z0$ pole will be summarised.
A comparison with measurements of $\as$ from other
processes will be given in Section~5, including an update of the
world summary of $\as$ determinations \cite{qcd96}. 

\section{QCD IN $\epem$ ANNIHILATION -- FROM SPEAR TO SLC AND LEP}

\begin{figure}
\vskip-20mm
\begin{center}
\epsfxsize6.0cm
\epsffile{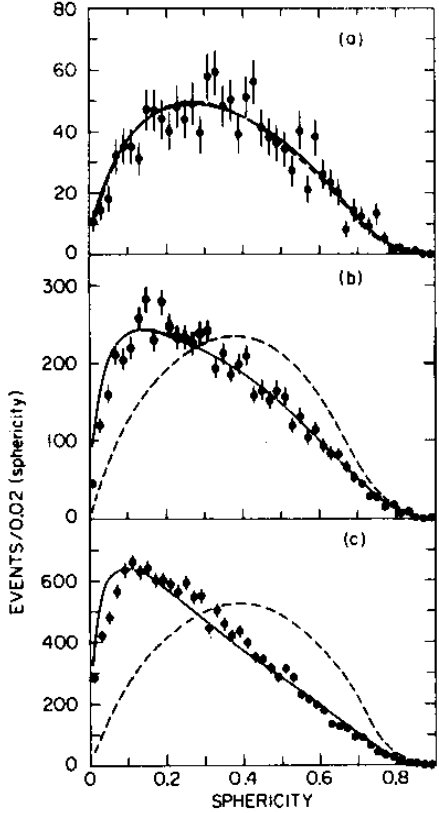}
\end{center}
\vskip-45mm
\caption{\label{2-jet}
Sphericity distributions observed at $\ecm = 3.0$~GeV~(a), 6.2~GeV~(b)
and at 7.4~GeV~(c), compared with a jet (solid curves) 
and a phase space model (dashed curves).
Data are from the SLAC-LBL Magnetic Detector at SPEAR 
\protect\cite{firstjets}.}
\end{figure}

\begin{figure}
\vskip-8mm
\epsfxsize7.5cm
\epsffile{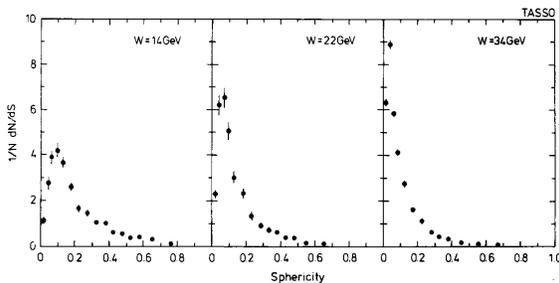}
\vskip-12mm
\caption{\label{tasso-s}
Sphericity distributions at $\ecm = 14$~GeV, 22~GeV
and 34~GeV (from TASSO \protect\cite{tasso-s}).}
\end{figure}

In 1975, the first \oq Evidence for Jet Structure in Hadron
Production by $\epem$ Annihilation" was reported \cite{firstjets}.
The data, taken with the SLAC-LBL magnetic detector at the SPEAR
storage ring, showed increasing evidence for the production of
two-jet like events when the center of mass energy, $\ecm$, was
raised from 3 to 7.4 GeV (see Fig.~\ref{2-jet}). 
The jet structure manifested itself as a decrease of the mean
sphericity, which is a measure of the global shape of hadronic
events.  
The angular distribution of the jet axes from the same measurement
provided evidence that the underlying partons must have 
spin~$\frac{1}{2}$. 
These observations, which were further corroborated
by  similar measurements
at $\ecm$ = 14 to 34 GeV at the $\epem$ storage ring PETRA
\cite{tasso-s}, see Fig.~\ref{tasso-s}, 
confirmed the basic ideas of the quark-parton model, 
which relates the constituents
of  the static quark model with partons produced in particle
collisions at high energies.

\begin{figure}[!htb]
\epsfxsize6.5cm
\epsffile{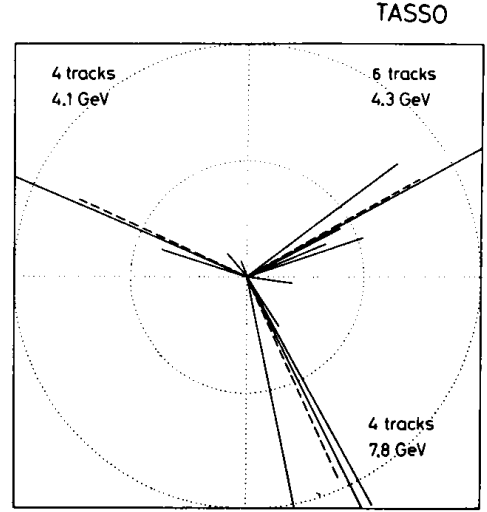}
\vskip-10mm
\caption{\label{tasso-3-jet}
The first 3-jet event observed by TASSO at PETRA.
Plotted are the momentum vectors of the charged particles, projected into the
principal event plane; the dotted lines indicate the reconstructed jet
axes (from \protect\cite{slwurep}).}
\end{figure}

\begin{figure}[!htb]
\epsfxsize6.5cm
\epsffile{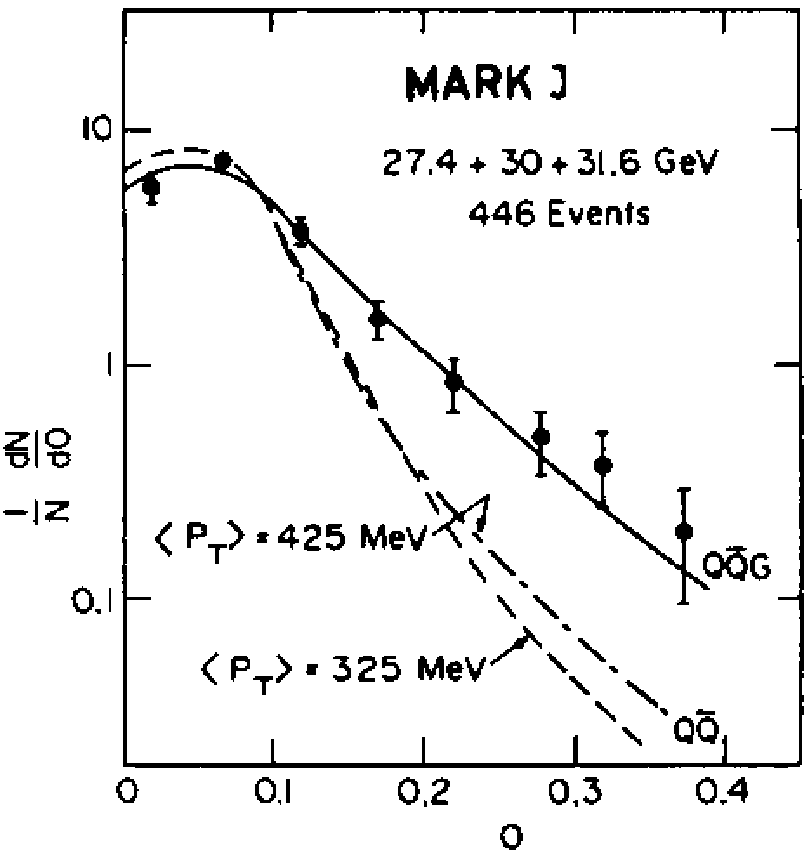}
\vskip-10mm
\caption{\label{markj-o}
Oblateness distribution measured at PETRA, compared with the
predictions based on a \protect $q \overline{q} g$ model (full line) and a
$q\overline{q}$ model (from \protect \cite{slwurep}).}
\end{figure}

\begin{figure}[!htb]
\vskip-5mm
\epsfxsize7.5cm
\epsffile{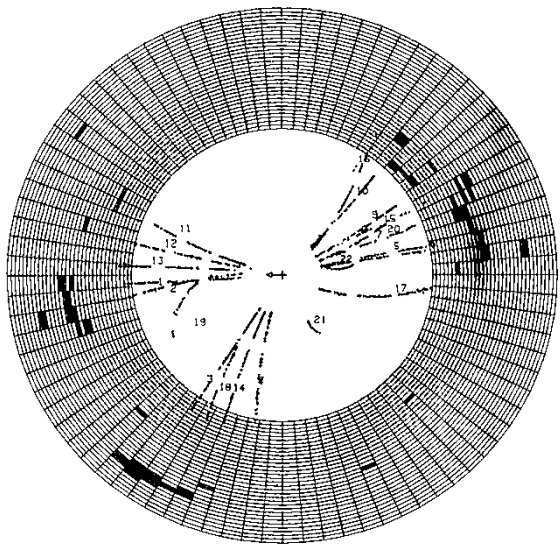}
\vskip-10mm
\caption{\label{Jade-3jet}
Detector view of a `typical' 3-jet event recorded with the JADE
detector at PETRA.}
\end{figure}

\begin{figure}[!htb]
\epsfxsize6.1cm
\epsffile{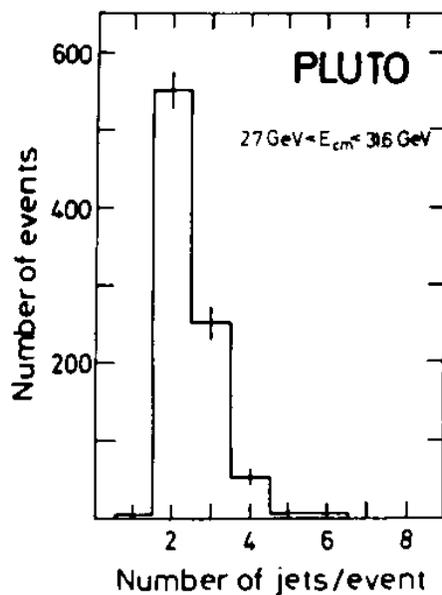}
\vskip-12mm
\caption{\label{pluto-njet}
First jet multiplicity distribution measured at PETRA \protect 
\cite{pluto-jets}.}
\end{figure}

In 1979, a small fraction of
planar 3-jet events was observed by the PETRA 
experiments around $\ecm \approx 30$ GeV \cite{gluon}, which
was attributed to the emission of a third parton
with  zero electric charge and spin 1 \cite{gspin}, as
expected for gluon bremsstrahlung predicted by QCD~\cite{egr}.
First evidence for 3-jet like
events came from visual scans of the energy flow within hadronic
events (Fig.~\ref{tasso-3-jet}; see also Fig.~\ref{Jade-3jet}), 
followed by statistical analyses of global
event shapes like oblateness (Fig.~\ref{markj-o}) and of 
angular correlations which are sensitive to the gluon spin
\cite{gspin}.
At the end of 1979, the MARK-J collaboration reported a first
measurement of $\as$, in first order perturbative QCD, from the
shape of the  oblateness distribution \cite{as-mkj}. 
The first analysis based on the definition and reconstruction of
resolvable jets was published in 1980 by the PLUTO collaboration
\cite{pluto-jets}, see Fig.~\ref{pluto-njet}. 

\begin{figure}[!htb]
\vskip-5mm
\epsfxsize7.5cm
\epsffile{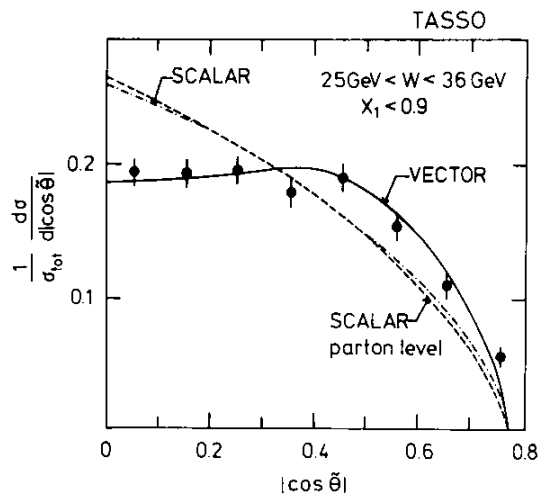}
\vskip-12mm
\caption{\label{tasso-gspin}
Distribution of the Ellis-Karliner angle \protect $\tilde{\Theta}$ measured by
TASSO \protect\cite{gspin}, compared with model predictions for a vector
and a scalar gluon.}
\end{figure}

In 1981, the JADE collaboration found  first evidence
\cite{jade-string} that the string hadroni\-zation model  
\cite{stringfrag} provides a better description of the
observed hadron flow in 3-jet events than does an independent jet
hadroni\-zation model. 

\begin{figure}[!htb]
\epsfxsize6.5cm
\epsffile{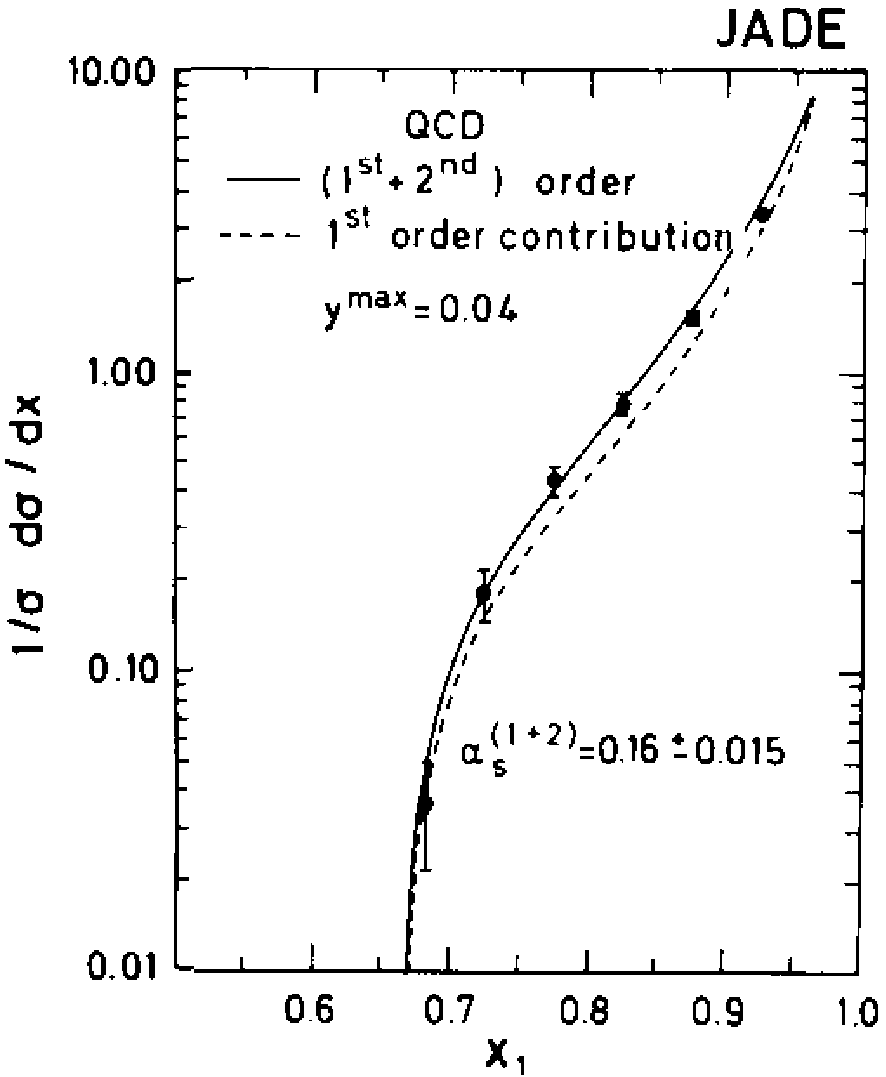}
\vskip-10mm
\caption{\label{jade-as2}
Corrected distribution of the scaled energy of the most energetic jet
of reconstructed 3-jet events, compared with analytic QCD predictions
in leading (dashed) and in next-to-leading order (full line) 
\protect \cite{jade-as2}.}
\end{figure}

\begin{figure}[!htb]
\vskip-4mm
\epsfxsize7.5cm
\epsffile{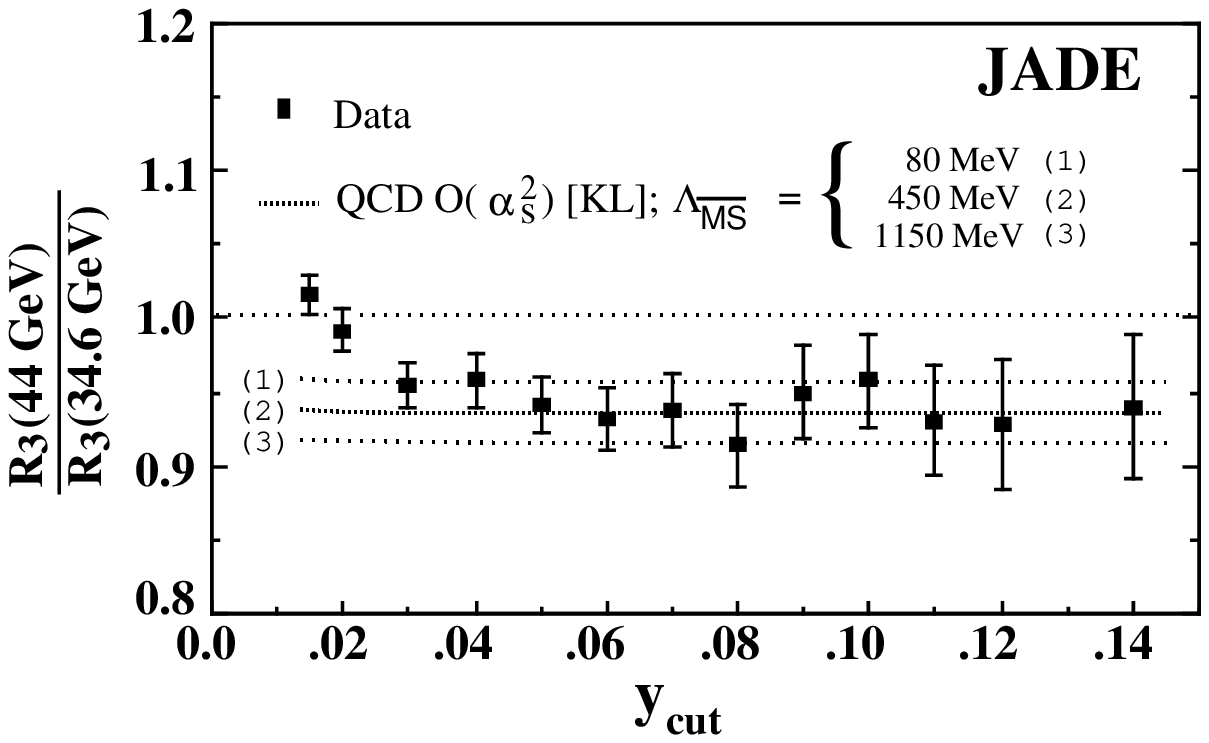}
\vskip-10mm
\caption{\label{jade-r3}
The ratio of 3-jet productions rates at 44 and at 34.6 GeV c.m. energy,
analysed with the JADE jet finder 
\protect \cite{jadejet2}.
The data are compared with QCD analytic predictions.
For the hypothesis of an $\as$ which does $not$ run with energy, a
constant ratio of 1 would be expected.}
\end{figure}

The year 1982 brought first evidence for 4-jet like events, observed
by JADE at $\ecm$ = 33 GeV \cite{jade-4-jet}, and the first
determination of $\as$ in second order perturbation theory ($\oaa$)~\cite{ert},
again by JADE \cite{jade-as2}.
The latter analysis was already
based on principles and ideas which nowadays are standard for most
of the $\as$ analyses at SLC and LEP:
An analytic QCD calculation in $\oaa$ perturbation theory was
fitted to a measured event shape distribution which had been
corrected for detector resolution and hadronization effects
(Fig.~\ref{jade-as2}).

\begin{figure}[!htb]
\epsfxsize7.5cm
\epsffile{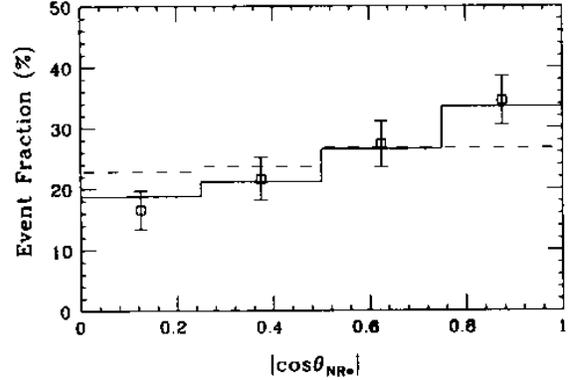}
\vskip-10mm
\caption{\label{amy-nr}
Distribution of the cosine of the modified Nachtmann-Reiter angle 
for 4-jet events measured by AMY at TRISTRAN
\protect \cite{amyjet}, compared with the predictions of a QCD (solid line) and
an Abelian (dashed line) model.}
\end{figure}

In 1986 JADE published the first detailed analysis of n-jet event
production rates \cite{jadejet1}, introducing a jet finding
mechanism which has since then been used in many other  studies.
The ratio of 4-jet over 3-jet event
production rates was found to be significantly larger than predicted
by $\oaa$ QCD, an observation that motivated studies of the
influence of the choice of renormalization scales in finite order 
perturbative QCD 
\cite{scales}. 
In 1988 it was
demonstrated by JADE that the energy dependence of 3-jet event
production rates gives evidence for the running of $\as$
\cite{jadejet2},
see Fig.~\ref{jade-r3}. 
First signs of the presence of the gluon self coupling
were observed in a study of 4-jet events by AMY \cite{amyjet}
around $\ecm$ = 56 GeV (Fig.~\ref{amy-nr}).

\begin{figure}[htb]
\epsfxsize7.5cm
\epsffile{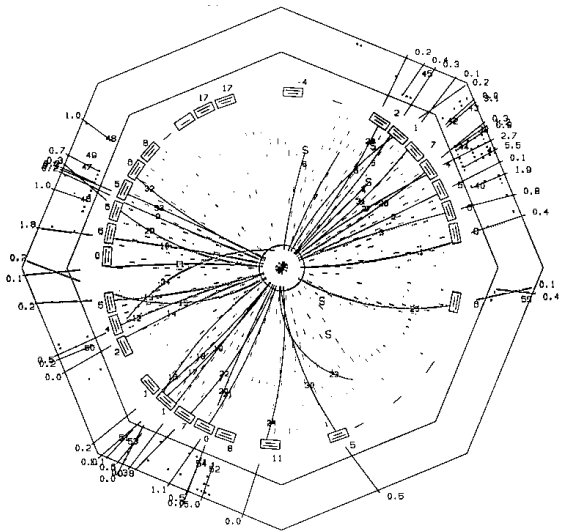}
\vskip-0.7cm
\caption{\label{markii-3jet}
One of
the first hadronic $\z0$-decays, observed with the Mark-II detector 
at the SLAC Linear Collider.}
\end{figure}
\baselineskip=12.0pt

More detailed reviews about jet
physics in $\epem$ annihilations below the $Z^0$ resonance
can be found e.g. in
\cite{slwurep,naroska,budapest} .

\section{QCD AT AND ABOVE THE $\z0$ POLE}

In 1989, both the SLC and LEP started to deliver 
$\epem$ collisions at and around the $\z0$ resonance, $\ecm \sim 91$~GeV.
An example of an event of the type $\epem \rightarrow Z^0
\rightarrow  3\ jets$, observed with the Mark-II detector at the SLC, is
shown in Fig.~\ref{markii-3jet}.

There exist several detailed reviews of QCD studies at LEP and SLC,
see e.g. \cite{annrev,hebbeker,scotland,burrows},
and many of the most recent results were presented
at this conference.
Here, only an overview of some of the major
achievements will be given:

\par \noindent 
New QCD calculations...

\begin{itemize}
\item
$\oaaa$ QCD predictions for the hadronic branching fraction of the
$\z0$ boson and the $\tau$ lepton \cite{gorishny,braaten},
\item
resummation of leading and next-to-leading logarithms and their
matching with caluclations in complete 
$\oaa$, for several hadronic event shapes
observables \cite{catani},
\item
$\oaa$ predictions for massive quarks \cite{wbquarks};
\end{itemize}

\noindent
new observables...
\begin{itemize}
\item
jet finding algorithms \cite{bkss}
like the Durham scheme \cite{durhamjets} or the cone algorithm \cite{conejets},
\item
event shape variables like total and wide jet broadening, $B_T$ and $B_w$
\cite{jetbroad},
\item
observables to study 4-jet kinematics \cite{d-a34};
\end{itemize}

\noindent
and new experimental techniques...
\begin{itemize}
\item
simultaneous use of many observables to study and reduce systematic
and theoretical uncertainties in measurements of $\as$ \cite{lepas2},
\item
variations of renormalisation scale, parton virtuality, quark masses
etc. to estimate theoretical uncertainties \cite{lepas2},
\item
tagging of quark - and gluon-jets \cite{qgtag},
\item
detailed studies of $\tau$-lepton decays (\cite{settles});
\end{itemize}

\noindent
... resulted in:
\begin{itemize}
\item  
precise determinations of $\as$ (see Sections~4 and~5),
\item
tests of the QCD group structure (see \cite{dissertori}),
\item
detailed studies of differences between quark- and gluon-jets
(see \cite{qgdiff}),
\item 
many details of the hadronisation process (see e.g. \cite{lafferty}),
\item
measurements of the heavy quark masses, $m_c(M_\tau)$ and $m_b(M_z)$
\cite{qmasses}.
\end{itemize}

\section{$\as$ AT LEP ABOVE THE $\z0$-POLE}

\begin{table*}[htb]
Table 1.\ \
Summary of measurements of $\as$ at LEP-1.5 and at LEP-2.\\
\begin{tabular}{|l|l|l|l|c|}
   \hline
 Exp. & $\as (\sqrt{s} = 133$~GeV) & $\as (\sqrt{s} = 161$~GeV) &
  $\as (\sqrt{s} = 172$~GeV) & refs.\\
\hline \hline
 ALEPH & $0.115 \pm 0.008 \pm 0.005$ & $0.111 \pm 0.009 \pm 0.005$ &
  $0.105 \pm 0.010 \pm 0.004$ & \cite{a-133} \\
 DELPHI & $0.116 \pm 0.007^{\ +0.005}_{\ -0.004}$ & $0.107 \pm
  0.008^{\ +0.005}_{\ -0.004}$ & $0.104 \pm 0.013^{\ +0.005}_{\ -0.004}$ &
  \cite{d-133} \\
 L3 & $0.107 \pm 0.005 \pm 0.006$ & $0.103 \pm 0.005 \pm 0.005$ &
  $0.104 \pm 0.006 \pm 0.005$ & \cite{l-as} \\
 OPAL & $0.110 \pm 0.009 \pm 0.009$ & $0.101 \pm 0.009 \pm 0.009$ &
  $0.093 \pm 0.009 \pm 0.009$ & \cite{o-133} \\
\hline
 Average & $0.111 \pm 0.003 \pm 0.007$ & $0.105 \pm 0.004 \pm 0.006$ &
  $0.102 \pm 0.004 \pm 0.006$ & \\
\hline
\end{tabular}
\end{table*}

The most recent data from LEP
around $\ecm$~=~133~GeV, 161~GeV and 172~GeV - although with rather
limited statistics of only a few hundred hadronic events for each
experiment and at each energy point - were analysed in terms of 
hadronic event shape distributions, jet production rates, charged
particle multiplicities, momentum distributions of charged particles
and other observables.
These analyses revealed that the new data are well described by
standard QCD and hadronisation models, which were tuned to the
high statistics data sets at the $\z0$ pole.
They also provided first measurements of $\as$ at these new energies which
are summarised in Table~1.

It should be noted that several of these measurements are not yet officially
published; those results as well as the combined values of $\as$ shown
in Table~1 are therefore still preliminary.


\section{UPDATE OF THE WORLD SUMMARY OF $\as$}

Significant determinations of $\as$, based on perturbative QCD
predictions which were, at least, complete to next-to-leading
order (NLO), date back to 1979/1980 (from structure functions in deep
inelastic scattering; see e.g. \cite{yndurain}) and to 1982 (from
$\epem$ annihilation \cite{jade-as2}).
Many of the early data and determinations of $\as$, see
e.g. \cite{yndurain-brighton}, have been superseded and/or replaced by
more actual measurements and analyses, c.f. 
\cite{altarelli-ac,virchaux,dallas,fernandez}.

\begin{table*}[htb]
Table 2. \ \
World summary of measurements of $\as$.
Entries preceded by $\bullet$ 
are new or updated since summer 1996~\cite{qcd96}.
Abbreviations:
DIS = deep inelastic scattering; GLS-SR = Gross-Llewellyn-Smith sum rules;
Bj-SR = Bjorken sum rules;
(N)NLO = (next-)next-to-leading order perturbation theory;
LGT = lattice gauge theory;
resum. = resummed next-to-leading order.
\begin{center}
\begin{tabular}{|r l|c|l|l|c c|c|}
   \hline
 & &  Q & & &  \multicolumn{2}{c|}
{$\Delta \amz $} &  \\ 
 & Process & [GeV] & $\alpha_s(Q)$ &
  $ \amz$ & exp. & theor. & Theory \\
\hline \hline \normalsize
$\bullet$ & DIS [pol. strct. fctn.] & 0.7 - 8 & & $0.120\ ^{+\ 0.010}
  _{-\ 0.008}$ & $^{+0.004}_{-0.005}$ & $^{+0.009}_{-0.006}$ & NLO \\
& DIS [Bj-SR] & 1.58
  & $0.375\ ^{+\ 0.062}_{-\ 0.081}$ & $0.122\ ^{+\ 0.005}_{-\ 0.009}$ & 
  -- & -- & NNLO \\
& DIS [GLS-SR] & 1.73
  & $0.32\pm 0.05$ & $0.115\pm 0.006$ & $ 0.005 $ & $ 0.003$ & NNLO \\
& $\tau$-decays 
  & 1.78 & $0.330 \pm 0.030$ & $0.119 \pm 0.004$
  & 0.001 &  0.004 & NNLO \\
$\bullet$ & DIS [$\nu$; ${\rm F_2\ and\ F_3}$]  & 5.0
  & $0.215 \pm 0.016$
   & $0.119\pm 0.005$   &
    $ 0.002 $ & $ 0.004$ & NLO \\
& DIS [$\mu$; ${\rm F_2}$]
     & 7.1 & $0.180 \pm 0.014$ & $0.113 \pm 0.005$ & $ 0.003$ &
     $ 0.004$ & NLO \\
& DIS [HERA; ${\rm F_2}$]
     & 2 - 10 &  & $0.120 \pm 0.010$ & $ 0.005$ &
     $ 0.009$ & NLO \\
& DIS [HERA; jets]
     & 10 - 60 &  & $0.120 \pm 0.009$ & $ 0.005$ &
     $ 0.007$ & NLO \\
$\bullet$ & DIS [HERA; ev.shps.]
     & 7 - 100 &  & $0.118\ ^{+\ 0.007}_{-\ 0.006}$ & $ 0.001$ &
     $^{+0.007}_{-0.006}$ & NLO \\
& ${\rm Q\overline{Q}}$ states
     & 4.1 & $0.223 \pm 0.009$ & $0.117 \pm 0.003 $ & 0.000 & 0.003
     & LGT \\
& $J/\Psi + \Upsilon$ decays
     & 10.0 & $0.167\ ^{+\ 0.015\ }_{-\ 0.011\ }$ & $0.113\ ^{+\ 0.007\ }
     _{-\ 0.005\ }$ & 0.001 & $^{+\ 0.007}_{-\ 0.005}$ & NLO \\
$\bullet$ & $\epem$ [ev. shapes]  & 22.0 & $0.161\ ^{+\ 0.016}_{-\ 0.011}$ &
   $0.124\ ^{+\ 0.009}_{-\ 0.006}$ & $\pm 0.005$ & $^{+0.008}_{-0.003}$
   & resum \\
& $\epem$ [$\sigma_{\rm had}$]  & 34.0 &
 $0.146\ ^{+\ 0.031}_{-\ 0.026}$ &
   $0.124\ ^{+\ 0.021}_{-\ 0.019}$ & $^{+\ 0.021}_{-\ 0.019}
   $ & -- & NLO \\
$\bullet$ & $\epem$ [ev. shapes]  & 35.0 & $ 0.143\ ^{+\ 0.011}_{-\ 0.007}$ &
   $0.122\ ^{+\ 0.008}_{-\ 0.006}$ & $\pm 0.002$ & $^{+0.008}_{-0.005}$
   & resum \\
$\bullet$ & $\epem$ [ev. shapes]  & 44.0 & $ 0.137\ ^{+\ 0.010}_{-\ 0.007}$ &
   $0.122\ ^{+\ 0.008}_{-\ 0.006}$ & $\pm 0.003$ & $^{+0.007}_{-0.005}$
   & resum \\
& $\epem$ [ev. shapes]  & 58.0 & $0.132\pm 0.008$ &
   $0.123 \pm 0.007$ & 0.003 & 0.007 & resum \\
& $\p\bar{\p} \rightarrow {\rm b\bar{b}X}$
    & 20.0 & $0.145\ ^{+\ 0.018\ }_{-\ 0.019\ }$ & $0.113 \pm 0.011$ 
    & $^{+\ 0.007}_{-\ 0.006}$ & $^{+\ 0.008}_{-\ 0.009}$ & NLO \\
& ${\rm p\bar{p},\ pp \rightarrow \gamma X}$  & 24.2 & $0.137
 \ ^{+\ 0.017}_{-\ 0.014}$ &
  $0.112\ ^{+\ 0.012\ }_{-\ 0.008\ }$ & 0.006 &
  $^{+\ 0.010}_{-\ 0.005}$ & NLO \\
& ${\sigma (\rm p\bar{p} \rightarrow\  jets)}$  & 30 - 500 &  &
  $0.121\pm 0.009$ & 0.001 & 0.009 & NLO \\
$\bullet$ & $\epem$ [$\Gamma (\z0 \rightarrow {\rm had.})$]
    & 91.2 & $0.124\pm 0.005$ & 
$0.124\pm 0.005$ &
   $ 0.004$ & $0.003$ & NNLO \\
& $\epem$ [ev. shapes] &
    91.2 & $0.122 \pm 0.006$ & $0.122 \pm 0.006$ & $ 0.001$ & $
0.006$ & resum. \\
$\bullet$ & $\epem$ [ev. shapes]  & 133.0 & $0.111\pm 0.008$ &
   $0.117 \pm 0.008$ & 0.004 & 0.007 & resum \\
$\bullet$ & $\epem$ [ev. shapes]  & 161.0 & $0.105\pm 0.007$ &
   $0.114 \pm 0.008$ & 0.004 & 0.007 & resum \\
$\bullet$ & $\epem$ [ev. shapes]  & 172.0 & $0.102\pm 0.007$ &
   $0.111 \pm 0.008$ & 0.004 & 0.007 & resum \\
\hline
\end{tabular}
\end{center}
\end{table*}

\begin{figure}[htb]
\epsfxsize7.5cm
\epsffile{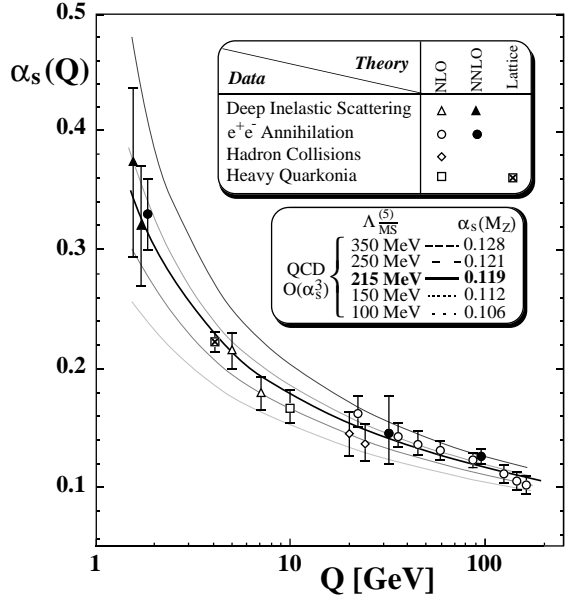}
\vskip-1.0cm
\caption{\label{asq97}
World summary of \protect $\as (Q)$ (status of July 1997).}
\end{figure}

\begin{figure}[htb]
\epsfxsize7.5cm
\epsffile{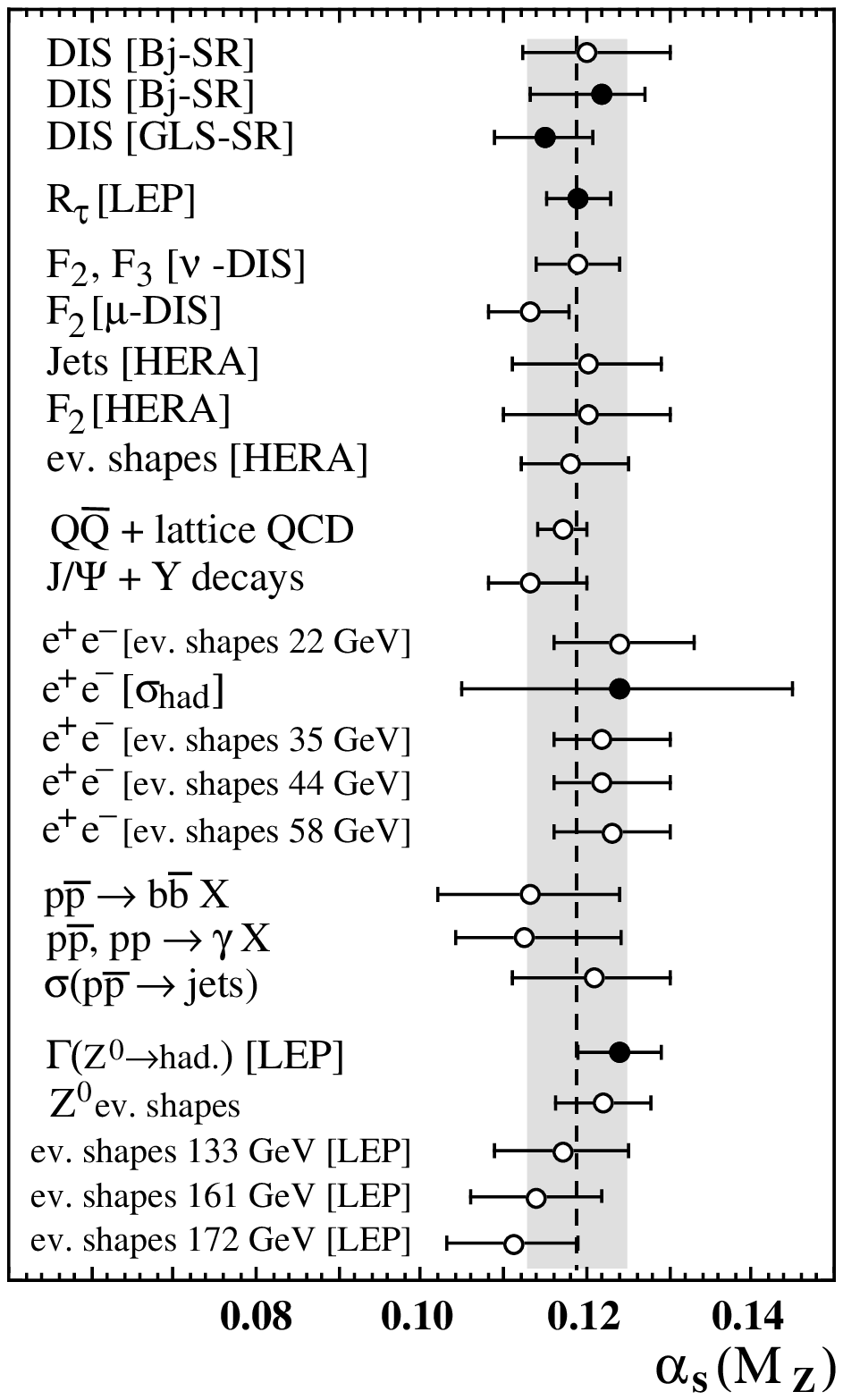}
\vskip-0.9cm
\caption{\label{asmz97}
World summary of $\amz$.}
\end{figure}

An update of a 1996 review of $\as$ \cite{qcd96} is given in Table~2.
Since last year, new results from polarized structure 
functions~\cite{altarelli-ball},
a correction of the results from $\nu$-nucleon deep
inelastic scattering \cite{ccfr}, new results from lattice QCD
\cite{davies}
and from a study of hadronic
event shapes at HERA \cite{h1-shapes}, a reanalysis of event shapes
from PETRA data \cite{fernandez}, an update of $\as$ from the
hadronic width of the $\z0$ \cite{lep-width} and new results from LEP
data above the $\z0$ pole (see Section~4) became available.
The data are displayed in Figures~\ref{asq97} and~\ref{asmz97}.

Perhaps the most relevant change of previously reported results is the
one from $\nu$-nucleon scattering \cite{ccfr}, which increased - due
to a new energy calibration of the detector - from $\amz = 0.111 \pm
0.006$ to $0.119 \pm 0.005$.
This partly resolves the outstanding problem that
deep inelastic scattering results preferred lower values of $\amz$ then did
measurements in $\epem$ annihilation.

\begin{figure}[htb]
\vskip-18.0mm
\epsfxsize8.5cm
\epsffile{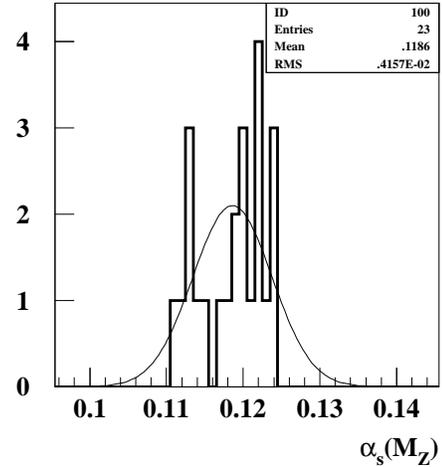}
\vskip-15.0mm
\caption{\label{asfreq}
Frequency distribution of central values of $\amz$ 
(histogram; data from Table~2) and a gaussian with the same mean and width
(curve).}
\end{figure}

\begin{figure}[htb]
\vskip-18mm
\epsfxsize8.5cm
\epsffile{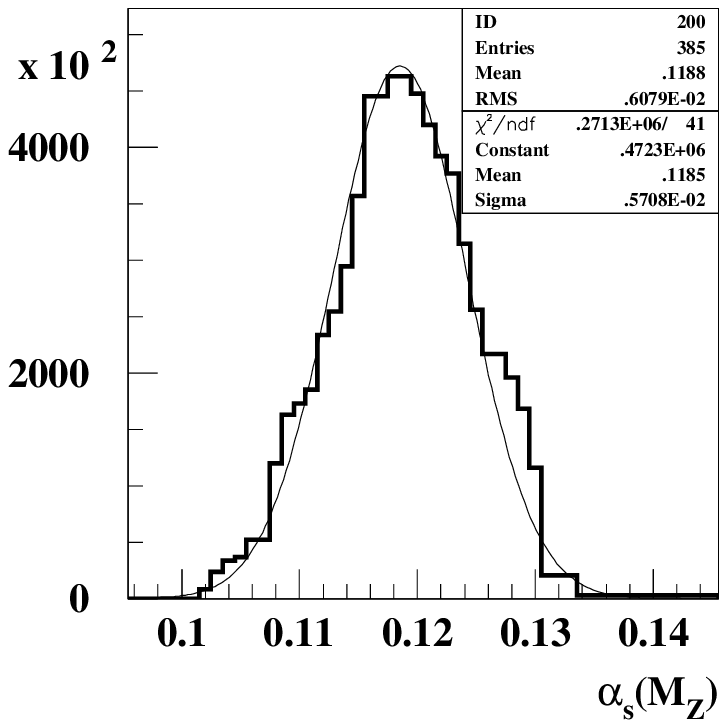}
\vskip-15.0mm
\caption{\label{asint}
Sum of weights of $\amz$ 
(histogram; calculated from data of Table~2) and
gaussian fit (curve).}
\end{figure}

The distribution of the central values of $\amz$ is
displayed in Figure~\ref{asfreq}, whereby the result from lattice QCD
is not included because the relevance of its given
uncertainty is not entirely clear.
The unweighted mean of this distribution is 0.1186, the root mean
squared (r.m.s.) is 0.0042.
This distribution, however, does not have a gaussian shape, and
so the r.m.s. is not regarded as a good measure of the error of the
average value of $\amz$.
Owing to the fact that the dominating uncertainties of most $\as$
measurements are theoretical rather than statistical or experimental,
which may then be correlated to an unknown degree,
two alternatives to estimate the uncertainty of the
average $\amz$ are applied:

First, it is demanded that at least 90\% of the central values of
all measurements must be inside the error interval. 
This gives
$\Delta \amz = \pm 0.006$, which can be regarded as a `90\%
confidence interval' (68\% confidence would correspond to $\pm 0.005$).

Second,
assuming that each result of $\amz$ has a boxed-shaped rather than a
gaussian probability
distribution over the interval of its quoted error, and summing
all resulting weights\footnote{The weight of a measurement is defined
to be the inverse of the square of its total
error.} for each numerical value of $\amz$, leads to the
distribution shown in Figure~\ref{asint}.
The (weighted) mean of this distribution is almost identical to the
unweighted average (see Fig.~\ref{asfreq}), $\amz = 0.1188$.
This distribution now has a gaussian shape; the
r.m.s. is 0.0061 and the width of a gaussian fit to that distribution
is 0.0057.
Therefore the world average result is quoted to be
$$ \amz = 0.119 \pm 0.006,$$
where the error of 0.006 is perhaps a conservative but well defined
estimate of the overall uncertainties.
This average corresponds to the full line in Figure~\ref{asq97} and to
the dashed line plus the shaded band in Figure~\ref{asmz97}.
The agreement between the different measurements is excellent. 
Due
to the many new and recent measurements, the running of $\as$ as
predicted by QCD is significantly proven by the data.

\end{document}